\documentclass[10pt]{article}
\usepackage[utf8]{inputenc}
\usepackage{authblk}

\usepackage{graphicx}
\usepackage{listings}

\usepackage{xcolor}
\usepackage[spaces,hyphens]{url}
\usepackage{hyperref}
\usepackage{breakurl}

\usepackage[strings]{underscore}

\newcommand{\cc}{C\&C }
\newcommand{\opret}{OP\_RETURN }

\begin{document}
\title{\vspace{-3.0cm}Leveraging Bitcoin Testnet for Bidirectional \\Botnet Command and Control Systems}
\date{}

\author[]{Federico Franzoni\thanks{The work of this author is partly supported by the Spanish Ministry of Economy and Competitiveness under the Maria de Maeztu Units of Excellence Programme (MDM-2015-0502).}}
\author[]{Ivan Abellan}
\author[]{Vanesa Daza\thanks{This author was supported by Project RTI2018-102112-B-I00 (AEI/FEDER,UE).}} %
\affil[ ]{Universitat Pompeu Fabra, Barcelona, Spain}
\affil[ ]{\texttt{\{federico.franzoni,vanesa.daza\}@upf.edu}}
\affil[ ]{\texttt{iabellan@pm.me}}

%
\maketitle              
\begin{abstract}
Over the past twenty years, the number of devices connected to the Internet grew exponentially.
Botnets benefited from this rise to increase their size and the magnitude of their attacks.
However, they still have a weak point in their Command \& Control (C\&C) system, which is often based on centralized services or require a complex infrastructure to keep operating without being taken down by authorities.
The recent spread of blockchain technologies may give botnets a powerful tool to make them very hard to disrupt.
Recent research showed how it is possible to embed \cc messages in Bitcoin transactions, making them nearly impossible to block.
Nevertheless, transactions have a cost and allow very limited amounts of data to be transmitted.
Because of that, only messages from the botmaster to the bots are sent via Bitcoin, while bots are assumed to communicate through external channels.
Furthermore, for the same reason, Bitcoin-based messages are sent in clear.
In this paper we show how, using Bitcoin Testnet, it is possible to overcome these limitations and implement a cost-free, bidirectional, and encrypted \cc channel between the botmaster and the bots.
We propose a communication protocol and analyze its viability in real life.
Our results show that this approach would enable a botmaster to build a robust and hard-to-disrupt \cc system at virtually no cost, thus representing a realistic threat for which countermeasures should be devised.

\end{abstract}

\section{Introduction}
A botnets is a network of infected devices, called \textit{bots}, collectively controlled by a single actor, called the \textit{botmaster}.
Botnets have been a major threat on the Internet for a long time, being used for a variety of malicious activities, like spamming, credentials stealing, and Distributed Denial of Service (DDoS) attacks\cite{liu2009botnet}.
A lot of research has been done to help detect and disrupt such activities on the web\cite{feily2009survey}.
However, the frequency and magnitude of botnet attacks drastically increased in the past few years, due to the massive adoption of computing devices and the advent of the Internet of Things (IoT), which is connecting millions of insecure devices to the web\cite{bertino2017botnets}.
Recent attacks from the infamous Mirai botnet\cite{antonakakis2017understanding}, showed the potential of this threat, with DDoS attacks of up to 1.1 Tbps\cite{kolias2017ddos}.

Meanwhile, blockchain is also becoming increasingly adopted as a tool for building distributed systems where different parties are able to exchange assets and data in a trustworthy manner\cite{nofer2017blockchain}.
Recent research showed how blockchains can be leveraged to implement the command and control (C\&C) system of a botnet\cite{mccorry2015zombiecoin}\cite{frkat2018chainchannels}.
In fact, using public blockchains, like Bitcoin, as the communication channel has several advantages for a botnet. 
First of all, they come with the strengths of all distributed networks, such as robustness and efficiency. 
Secondly, they are not regulated by any authority, making them censorship-resistant, meaning that no specific content or user can be banned.
Furthermore, they privilege privacy, by making use of pseudonyms and hindering the association between a transaction and the device that generated it.
As such, although possible\cite{biryukov2014deanonymisation}\cite{koshy2014analysis},
it is not trivial to identify nodes participating in a botnet, and even more importantly, to identify the botmaster.
All such properties are ideal for a botnet\cite{silva2013botnets}, as they allow to operate, protected, over a long period of time, with virtually no risk of having communications disrupted.

Most state-of-the-art research proposes Bitcoin transactions as the main \cc vector, following different strategies to embed commands from the botmaster.
However, these proposals have important limitations.
First of all, they only cover communications from the botmaster, delegating replies form the bots to external channels, typically employing a web server.
Furthermore, messages are very limited in size and are sent in clear, as cryptography is only implemented on the external channel.
Finally, messages have a cost, since they are sent via transactions.
All these limitations make this approach seem impractical or inconvenient for a real-world botnet implementation.

In this paper, we show it is possible to overcome such limitations by leveraging the Testnet network, instead of Mainnet.
We propose a bidirectional communication protocol that implements encryption and allows bigger amounts of data to be exchanged.
To the best of our knowledge, this is the first paper to study bidirectional \cc communications on top of Bitcoin.
Our approach makes a fully-blockchain-based botnet implementation both practical and economical.

\paragraph{Organization of the Paper}
The rest of the paper is organized as follows.
Section \ref{sec:background} describes the necessary background topics, that is Botnet \cc and Bitcoin.
Section \ref{sec:related} covers previous research work and discuss its limitations.
In Section \ref{sec:testnet}, we show the advantages of using Testnet as the \cc channel.
In Section \ref{sec:design}, we describe our communication protocol design and in Section \ref{sec:results} we show our experimental results.
Section \ref{sec:discussion} analyzes the viability and robustness of our proposal.
Section \ref{sec:conclusion} concludes the paper and discuss future work.

\section{Background}
\label{sec:background}
\subsection{Botnet \cc Communication}
In order for a botnet to operate, a communication channel is needed between botmaster and bots.
The infrastructure used for that purpose is known as the Command \& Control (C\&C) system.
This is a crucial component for a botnet, as it is the only means to keep control over the bots.
As such, it is has to be designed carefully, in order to avoid being disrupted.
In other words, the \cc system should allow controlling the botnet as long as possible, providing stealthy and efficient communication between botmaster and bots.

Strategies to implement \cc changed over the years, following the evolution of available technologies and the ability of authorities to counter existing approaches\cite{mahmoud2015survey}.
First-generation botnets leverage hardcoded Internet Relay Chat (\verb|IRC|) channels, where bots connect to receive instructions from the botmaster. 
This system is simple and cheap but is also easy to detect and take down\cite{binkley2006algorithm}\cite{aburajab2006multifaceted}.
Second-generation botnets make use of \verb|HTTP|, with hardcoded web domains, periodically contacted by the bots to download instructions. This approach allows to effectively blend messages into legitimate Internet traffic. Nonetheless, effective techniques exist to detect botnet communications\cite{livadas2006using}\cite{gu2008botsniffer}, allowing to quickly shutdown malicious domains\cite{westervelt2009botnet}.

Early botnets relied on a client-server model, thus having a central point of failure, which can always be detected and shut down by the authorities.
Last generation botnets overcome this issue by adopting a P2P model. 
Bots and \cc server connect as peers to the same network, making it difficult to distinguish the source of the commands\cite{wang2009systematic}. 
This architecture makes the botnet much more robust and hard to shut down.
Nonetheless, it is still possible to detect P2P-botnet traffic using advanced techniques\cite{nagaraja2010botgrep}\cite{saad2011detecting}. 
Moreover, to join the network, bots need hardcoded addresses, which can be easily blocked by authorities if detected.
Modern botnets tend to use a mix of techniques, such as P2P network with HTTP \cc server, or leverage cloud-based services and social media as rendezvous points\cite{mahmoud2015survey}. 
Although these services are easy to setup and access, providers can promptly block any detected malicious account.

\subsection{Bitcoin}
Bitcoin is a digital payment system released in 2009. Participating actors are identified by alphanumeric strings called \textit{addresses}. Each address represents the public part of an (asymmetric) cryptographic key pair, whose private part is used by the owner to sign transactions. When a coin is sent to a specific address, only the owner of the corresponding private key can spend it.
Transactions are validated by nodes of a P2P network that cooperate to maintain a distributed ledger, structured as a chain of blocks (or \textit{blockchain}). 
Each block contains a set of valid transactions and is linked to the previous one by including its hash.
Blocks are concurrently created by special nodes called \textit{miners}, which compute the solution of a cryptographic puzzle over the transactions of the new block.
This solution is known as \textit{Proof of Work} (PoW) and is included in the block itself.
Transactions and blocks are validated and distributed by all the peers of the network.
To decide on conflicting versions of the ledger, peers always choose the longest chain they know, that is the one with the biggest PoW. 
By following this scheme, the ledger is considered to be immutable and able to avoid \textit{double spending} the same coin\cite{nakamoto2008bitcoin}.

\paragraph{OP\_RETURN}
Since 2014, it is possible to embed a small amount of data inside a transaction, using the \opret opcode\cite{wiki2019opreturn}. This possibility was introduced to discourage other wasteful methods of embedding data, such as using non-existing transaction output addresses. The new opcode allows adding a non-spendable output, which carries up to 80 bytes of arbitrary data.
\opret is often used to implement asset exchange protocols on top of Bitcoin or to add valuable data in the blockchain\cite{bartoletti2017analysis}.

\paragraph{Testnet} 
As other public blockchains, Bitcoin provides a separate network for developers to test their applications, known as Testnet\cite{wiki2019testnet}.
While running the same protocol as the main network (Mainnet), Testnet has some important differences.
First of all, Testnet coins (tBTC) have no real value, and can be easily obtained via online services called \textit{faucets}. 
Secondly, the mining difficulty is also set to a lower value than Mainnet, making the blockchain grow faster.
Finally, some restrictions are ignored to allow developers to test edge cases. In particular, non-standard transactions are allowed, thus being relayed and mined by the network.
We will see how these and other characteristics significantly help implementing a botnet C\&C.

\paragraph{Bitcoin nodes}
There are two main options to access the Bitcoin blockchain: full nodes and Simple Payment Verification (SPV) nodes.
Full nodes are the building blocks of the P2P network.
They validate all transactions and blocks, and relaying them to their peers.
This is the most secure way to use Bitcoin, but requires to download the whole blockchain, which can be very resource-consuming.
SPV nodes, like full nodes, receive and relay all transactions, but do not download the whole blockchain.
Instead, they only download block headers and rely on other peers to retrieve the blocks they need to validate transactions of interest. 
This make the node suitable for resource-constrained devices at the expense of a certain level of trust into other peers.
Thanks to their better performances, SPV nodes are today the most popular choice on Bitcoin\cite{sheinix2018bitcoin}.

\section{Related Work}
\label{sec:related}
\textit{ZombieCoin}\cite{mccorry2015zombiecoin} was the first paper to propose Bitcoin as a means for \cc communications. Bots embed the botmaster public key and decode transactions coming from the corresponding address. To embed commands, the \opret opcode is used, which allows to carry up to 80 bytes of data.
In \cite{ali2018zombiecoin} the same authors propose enhancements such as \textit{transaction-chaining} to embed longer messages and external upstream communication by means of periodical \textit{rendezvous-point} announcements.
The main limitations of this proposal are the server-based upstream communication and the cost of messages sent on the blockchain.
The authors claim that it would be impractical and economically prohibitive to implement upstream communication on top of the blockchain. We show that this is not true when leveraging Testnet.

\textit{ChainChannels}\cite{frkat2018chainchannels} proposes a more generic approach, which can be used on different blockchains as it does not leverage Bitcoin-specific features. The authors describe a method to insert hidden data into transaction signature, which can be later decoded with the private key used for the signature. 
For this purpose, the authors propose a key-leakage scheme that allows bots to decipher messages at a later time.
This is a very portable approach, since virtually all blockchains employ digitally-signed transactions with a compatible signature scheme.
Nonetheless, this approach suffers from the same limitations as ZombieCoin: messages are costly and limited in size; communication is unidirectional and unencrypted.
Furthermore, bots can only decrypt messages in a second moment, assuming they execute commands altogether after these have been issued, something that might not be realistic.

In \cite{baden2019whispering}, the authors propose an approach based on \textit{Whisper}, a communication protocol that runs on top of the Ethereum network.
This approach does not use transactions and thus has no cost.
It also provides a good level of privacy and allows for two-way communication. Moreover, as messages are not in transactions, they are not added to the blockchain, making their backward analysis harder.
However, Whisper, which is still in a PoC stage, it is not enabled by default on the standard Ethereum client (geth) and there are no known statistics about how many nodes currently run the protocol. 
Consequently, its reliability is unknown, making it unlikely to be actually used by a botnet as of today.

\section{Leveraging the Testnet Network}
\label{sec:testnet}
As explained in Section \ref{sec:background}, Bitcoin Testnet follows the same protocol as the Mainnet but has some important differences.
In particular:
\begin{itemize}
	\item Testnet coins have no value in real life. For this reason, they can be easily obtained for free through online services called \textit{faucets}\cite{wiki2019testnet}.
	\item Mining is much easier, since the PoW difficulty is set to a lower value. As a consequence, unlike Mainnet, it is feasible to run a solo miner\cite{lopp2015how} to earn coins.
	\item The Testnet network and blockchain are about ten times smaller than Mainnet\cite{delgado2018txprobe}. 
	This makes clients synchronize faster and consume less resources.
	\item Non-standard transactions are validated and relayed by the network. This feature enables the following characteristics:
	\begin{itemize}
		\item \opret can be bigger than 80 bytes. In fact, there is no explicit limit to the amount of data that can be actually embedded;
		\item Transactions can have multiple outputs with the same address as well as multiple OP\_RETURNs;
		\item Transaction outputs can be below the dust limit\footnote{On Bitcoin, it is considered \textit{dust} any output smaller than the amount needed to spend that output. Hence its value actually depends on the transaction size, but its minimum is usually considered to be 546 satoshis};
		\item Transaction size can be greater than the maximum (which is around 100kB).
	\end{itemize}
\end{itemize}

All these properties give numerous benefits for the implementation of a botnet.
First of all, the botmaster can easily obtain the necessary amount of coins to run its botnet, either by using faucets or running a miner.
Secondly, the reduced size of Testnet blockchain and network make bots less resource-demanding, allowing them to hinder detection and even to run on low-resource devices.
Finally, non-standard transactions give the ability to send bigger and more complex messages.

These features allow overcoming all the main drawbacks of previous Bitcoin-based proposals:
botnet communications have no cost thanks to the fact that Testnet coins have no real value;
bidirectional communication can be implemented thanks to the great number of coins that can be obtained for free;
encryption can be implemented thanks to the larger amount of data that can be embedded in each transaction.

\section{Botnet Design}
\label{sec:design}
In this section, we propose a viable communication protocol for Testnet, based on non-standard transactions, that provides a bidirectional and encrypted \cc channel at zero cost.

As in previous works, we assume there exist an infection mechanism that takes control of devices and downloads the bot client. 
The botnet is composed by a \cc server node, directly controlled by the botmaster, and a number of bot nodes.
We assume the \cc server is not resource-constrained and runs a full node.
On the other side, bots run an SPV node to consume less resources and hinder detection.

In the rest of this section, we explain how the communication works (transactions, fees and encryption) and describe the different phases of the protocol (registration, commands and responses).

\subsection{Communication}
All communications between the botmaster and the bots happen through transactions.

\paragraph{Data Embedding and Fees}
We use \opret outputs to embed messages inside transactions.
As previously mentioned, this operator has no explicit limits of size on Testnet. 
As such, the amount of data that can be embedded is only limited by the maximum size of a transaction, which, again, is not explicitly limited on Testnet. This makes the theoretical size limit bound by the size of a block (around 1 MB). 
However, a practical limit to this amount is given by the minimum fee needed to have the transaction relayed by other peers.
This value is known as the \textit{minimum relay fee} (MRF).
MRF does not differ between Mainnet and Testnet and is proportional to the size of the transaction itself. 
This means that, although sending very large messages is possible, this can be excessively expensive in terms of fees. 
We will see more details about MRF later in Sections \ref{sec:results} and \ref{sec:discussion}.

In our protocol, all transactions spend a fee equivalent to the corresponding MRF.
To this respect, it is important to notice that using low fees might make the transaction mined later.
However, from the botnet perspective, it is not important if and when messages are added to the blockchain, but only if they travel across the network and reach the \cc server.

\paragraph{Encryption and Authentication}
In order to protect communications, we use encryption in both directions. 
To obtain the best compromise between security and efficiency, we make use of an hybrid approach.

We assume the botmaster creates an asymmetric key pair, called \textit{botmaster keys} before the creation of the botnet and hardcode bots with the public key. This key pair is completely unrelated to the address used to send commands, which in fact, can change at every message.
Additionally, a symmetric key is also embedded in the bots, called \textit{botnet key}.

For the sake of clarity, we distinguish between \textit{downlink} encryption, used from the botmaster to the bots, and \textit{uplink} encryption, used by the bots to communicate with the botmaster.

Downlink encryption works this way: when the botmaster wants to send a command, it encrypts it with the botnet key and signs it with its private key; when bots receive a transaction with an OP\_RETURN, they check the signature using the botmaster public key. If the signature is valid, they decrypt the message with the botnet key and execute the command.
This scheme allows the bots to recognize transactions from the botmaster even without knowing its address.
Moreover, thanks to the signature, bots are assured about the authenticity of the source.

For uplink encryption, each bot creates a private symmetric key, called the \textit{bot key}, which is sent to the botmaster at the time of registration, encrypted with the botmaster public key.
When sending messages, bots encrypt data with their bot key.
Furthermore, bots use a new address for each message, which corresponds to the change address of the previous transaction.
In order to recognize and decrypt bots messages, the botmaster keeps track of the current address of each bot and the corresponding encryption key. 

\paragraph{Transactions}
We have the following types of transactions: \textit{quotas}, \textit{registrations}, \textit{fundings}, and \textit{messages}.
Quotas have one input and several outputs (the quotas), which are used as input for the registration transactions.
Registration transactions have one input (a quota) and one \opret output. 
The quota equals the MRF for the registration message, so no change output is required.
Funding transactions have one input and one output, which equals the input value minus the MRF.
Messages (commands and responses) always have two outputs, one with the \opret carrying the message and the other sending the change (minus the MRF) to another address belonging to the sender (i.e. the \textit{change address}).

\subsection{Bot Registration}
When a new bot joins the network, the first thing it needs is to get some funds to send transactions.
As the bot cannot obtain funds autonomously (like the botmaster does), it needs to ask the botmaster to provide some.
However, at the same time, the botmaster needs to know the address of the bot in order to send such funds.

We solve this problem by having all bots sharing a common private key, that gives access to all transactions of an address called the \textit{shared account}.
The botmaster periodically puts funds on the account, while new bots use such funds to register to the botnet.
They do so by sending a \textit{registration message} which contains their own address and encryption key.
Since SPV clients do not store the UTXO set (the set of unspent transactions), they ask their peers about any available fund on the account.
The botmaster monitors transactions sent from the shared account and when it detects one, it stores the information about the new bot and sends it some funds.
After the registration, bots will only receive funds directly from the botmaster.

If more bots try to register at the same time, there might be a conflict between their transactions (i.e. a double spend). In order to minimize this risk, the botmaster puts on the account several transactions, called \textit{quotas}, containing just the right amount of coins needed to send the registration message.
Furthermore, to reduce concurrency, it always sends multiple quotas at the same time.
When a new bot wants to register, it picks a random quota and tries to send the message.
It then sets a timeout for receiving the funding from the botmaster.
If the timeout expires, the bot picks another quota and repeats the process.
The same happens if its transaction gets rejected by peers or if another transaction spending the same quota is detected.
At any time, the botmaster makes sure there are enough quotas on the shared account, according to the rate at which new bots are joining. 

Since the registration transaction comes from a shared account and only has an \opret output, neither the botmaster address nor the bot one are revealed.

It is worth noting that creating quotas would not be possible on Mainnet, as they would be considered as dust outputs and rejected by the network.

\subsection{Commands and Responses}
We distinguish between commands, that are messages sent by the botmaster, and responses, that bots send after executing a command.
Bots can execute three types of commands: \textit{hardcoded}, \textit{shell} and \textit{script}.

Hardcoded commands are functions that are already implemented by the bot code.
They can be executed once or repeated over a period of time.
Examples of hardcoded commands include a DoS function to attack a target or a keylogger to steal credentials.
The botmaster can send parameters such as interval and number of iterations, or make the function run indefinitely until it sends a stop command.

Shell commands are command-line instructions that the bot directly execute on the infected machine.
When the bot receives such command, it runs it and converts the output into a hexadecimal string to be sent as a response.

Script commands work similarly, but they use code stored on the blockchain.
In particular, the code to execute is embedded by the botmaster in a previous transaction, called \textit{script transaction}, and encrypted with a symmetric key, which is unknown to the bots.
The command includes the transaction ID of the script transaction and the key to decrypt.
When bots receive these commands, they retrieve the data, decrypt the payload and execute the code.
They then convert the output into a hexadecimal string and send it the botmaster.
In order to ensure all bots can send their response, the botmaster checks current funds of each bot before sending the command. If any bot does not have sufficient funds, the botmaster sends them more coins.

This approach takes advantage of the larger storage capacity of transactions on Testnet, 
which allow storing kilobytes of code on the blockchain.
Additionally, this technique enables the botmaster to reuse the same code several times, saving coins and reducing its traffic. 
By using shell and script commands, bots are not limited to the functions their code implements, but are able to perform a variety of attacks, making it harder to estimate their real capacity.

\section{Experimental Results}
\label{sec:results}
We created a PoC botnet that implements our protocol, and then, we simulated its basic activities.
In particular, we verified the ability to send, receive, execute and reply to commands.
We then calculated the necessary amounts of coins needed for each type of transaction we use.
Our results show that the proposed protocol is both viable and sustainable.

\subsection{Non-standard transactions and fees}
As a preliminary step, we verified the ability to send non-standard transactions on the network.
We also tested the limits we could reach while still having transactions relayed.

As stated in Section \ref{sec:testnet}, non-standard transactions allow us to do the following:
\begin{itemize}
	\item send \opret outputs that are larger than 80 bytes,
	\item send repeated outputs, both \opret and addresses,
	\item send dust outputs,
	\item send transactions larger than 100 kB.
\end{itemize}

We used Bitcoin Core v0.18.0 to perform our tests. We had to patch its code to allow creating transactions with repeated outputs (\opret or address).
All other tests were possible without any modification.

For what concerns \opret size, we successfully sent transactions carrying as much as 50 kB of data. 
All transactions got immediately relayed and, after some time, mined.
Although theoretically possible to send more, we were not able to send transactions carrying more data due to a limitation on the size of the argument that can be passed through the Linux command line\footnote{This is a known limitation of the Linux kernel; the actual argument size limit depends on the stack size of the system\cite{mascheck2016argmax}.}. 
As such we were not able to verify the ability to send transactions bigger than 100kB. However, we are confident this is actually possible, as this limit is not enforced for non-standard transactions.

Transactions with repeated outputs, both addresses and OP\_RETURN, were also accepted and relayed by all peers.

For what concerns dust outputs, we successfully sent transactions with as little as 0 satoshis, having them relayed and mined.

\subsection{PoC}
We implemented the \cc server with our patched version of Bitcoin Core, while bots run an SPV node using \verb|bitcoinj|, which did not need any modification to use our protocol.
Both bots and the \cc server run on a Linux operating system.

\paragraph{Encryption}
For asymmetric encryption and digital signature, we use RSA with a 2048-bit key and OAEP padding, which generates outputs of 256 bytes.
This allows bots to send up to 214 bytes of encrypted data to the botmaster.

For symmetric encryption we use AES with 256-bit keys, using CRC block mode and PKCS5 padding.
This encryption mode requires a random 128-bit IV (Initialization Vector), which is also needed for decryption.
As the IV does not need to be secret, we send it in clear along with the cyphertext.

\paragraph{Fees}
The default MRF value on Bitcoin Core clients is set to a value of 1000 satoshis (sats) per kB.
However, with the introduction of the so-called Segregated Witness (BIP141), transaction fees became dependent on what is known as virtual size, which is a function of the actual transaction size
\footnote{The virtual size \emph{v} is computed as \emph{v=(w+3*s)/4}, where \emph{w} is the size of the transaction and \emph{s} is the size of the corresponding base transaction (without the witness). In case of non-SegWit transaction, the virtual size is the actual size.}.
More specifically, the current MRF is calculated as 1 sat/vB, where vB stands for \textit{virtual Byte}.

In our implementation, we make use of the embedded functions of the clients to calculate this value for each transaction.

\paragraph{Transactions}
As stated in Section \ref{sec:design}, we have the following types of transaction: quotas, registration, fundings, commands and responses.
All transactions in our protocol have only one input.

Quotas transactions have 11 outputs, corresponding to batches of 10 quotas plus the change address.
Each quota corresponds to the MRF of a registration message.

Registration messages have a quota as the input and 1 \opret output containing the payload.
The payload contains a 36-byte-long Testnet address and a 32-byte-long AES key, encrypted with the public RSA key of the botmaster, which generates an output of 256 bytes. 

Fundings contain two outputs: the bot address, receiving the funds, and the change address of the botmaster.

Commands and responses have 1 \opret output, plus the change address of the sender.
Hardcoded commands have 3 bytes for the command plus the arguments (e.g. a target).
The payload is encrypted with AES, so their output size corresponds to the size of the payload, padded to fit the block size (16 bytes), plus the IV (16 bytes).
So, for example, an instruction like \verb|dos www.domain.com|, which is 19-byte long, will have a data output of 32 bytes.
Adding the IV we have 48 bytes. 
The script command has the following format: \verb|scr TXID KEY|, where TXID is a 32-byte-long transaction ID and the key is a 32-byte AES key. The corresponding IV is stored alongside the script itself.

\paragraph{Commands}
We implemented the following commands:
\textit{dos} and \textit{stop} as hardcoded commands, \textit{lshw} as shell command, and one script command called \textit{screenshot}.
After executing shell and script commands, bots convert the output to a hex string and send it as a response message.
To convert outputs into hex they use the following command: \verb!$(CMD) | tr -d '\n' | xxd -r -p!, where \verb|CMD| stands for the command they are executing.
The \verb|dos| command makes the bot attack a specific target, which is sent as a parameter.
The DoS attack is performed using \verb|hping3| and can be interrupted by a \verb|stop| command.
This command has no output.
The \verb|lshw| shell instruction makes the bot gather information about the hardware of the infected machine.
On our bot machine, this command generates approximately 12 kB of data.
The \textit{screenshot} script is shown in Listing \ref{lst:screenshot}.

\lstset{
	basicstyle=\footnotesize, frame=tb,
	xleftmargin=.1\textwidth, xrightmargin=.1\textwidth
}
\begin{lstlisting}[float=b, frame=single, caption={The \textit{screenshot} script},captionpos=b, label={lst:screenshot}] 
import -window root screenshot.png
convert -quality 5 screenshot.png screenshot.jpg
cat screenshot.jpg
\end{lstlisting}

This script takes a screenshot in PNG format, which is around 500 kB, then compress it to JPEG format, reducing its quality to fit into 50 kB of data. The \verb|cat| command dumps the content of the file to produce the output to send as a response.

\section{Discussion}
\label{sec:discussion}
In this section, we analyze the sustainability of our protocol in terms of the amount of coins needed to run a botnet, as well as the robustness of its architecture and the security of its design.

\subsection{Cost Analysis}
\paragraph{Funding the botnet}
At the time of writing, we were able to find six active faucets on the web.
The amount of coins obtained per request varies from 0.0001 to 0.089 tBTC, with an average of 0.05 tBTC per request. 
By making a single request per faucet, we obtained approximately 0.12 tBTC. 
Requests are usually limited by faucets to one per day, for each given IP address.
However, it is not hard to bypass the limit by using VPNs or proxy services.
Furthermore, as previously discussed, a botmaster could run a miner to obtain a much greater amount of coins, without any restriction.

As such, we consider the estimate of 0.1 tBTC per day as a conservative lower bound of the funds that a botmaster can obtain to operate its botnet.
In a real-life context, it is likely feasible to obtain ten to hundred times more than such an amount.

\paragraph{Protocol messages cost}
As discussed in Section \ref{sec:design}, all messages sent by the botnet spend the minimum relay fee (MRF), which is directly proportional to the size of the message and calculated as 1 satoshi per virtual byte. 

In our protocol, transactions can have a fixed size, like quotas, registrations, and fundings, or variable size, like commands, responses, and scripts.
Table \ref{tab:messagefees} shows the MRFs for all transactions used in our protocol.
For a quota batch transaction, which has 11 outputs, a MRF of 454 sats is needed.
Registration transactions have a payload of 256-byte long, corresponding to a MRF of 373 sats.
Fundings, which have 2 outputs, can be sent with 166 sats.
Commands payload size is the smallest multiple of the AES block size (16 bytes), plus the IV (16 bytes).
To simplify things, we assume hardcoded commands are short enough to fit into 2 blocks (32 bytes), which adds up to 48 bytes, with the addition of the IV. To send such a transaction, a fee of 161 sats is needed.
We also assume shell commands are smaller than 100 bytes, with bigger instruction sent as scripts.
Since the minimum size is 17 bytes (1-byte command plus the IV), the MRF varies from 133 to 230 sats.
Script commands have a 3-byte command plus a 32-byte transaction ID, a 32-byte script encryption key and the IV.
This sums up to 83 bytes, requiring a MRF of 197 sat.
We assume the maximum size of script transactions and responses is 50kB. 
For what concerns our non-hardcoded commands, we have the following values.
The encrypted screenshot script, along with the IV, is 128-byte long, corresponding to a MRF of 242 sats.
To send the response (50kB), 51349 sats were needed.
To send the output of \verb|lswh| (12kB), 12860 sats were needed.

\begin{table}[]
	\centering
	\begin{tabular}{|c|c|c|}
		\hline
		\textbf{Message}              & \textbf{\opret (Bytes)} & \textbf{Fee (Satoshis)} \\ \hline
		Quotas Batch         & N/A                             & 454            \\ \hline
		Registration (quota) & 256                           & 373            \\ \hline
		Funding              & N/A                             & 166            \\ \hline
		Hardcoded Command    & 48                            & 161            \\ \hline
		Shell Command        & 17 - 116                        & 133 - 230      \\ \hline
		Script Command       & 83                            & 197            \\ \hline
		Script (Transaction) & 117 - 51200                     & 231 - 51349    \\ \hline
		Response             & 17 - 51200                      & 133 - 51349    \\ \hline
	\end{tabular}
	\caption{Minimum relay fees for our protocol transactions}
	\label{tab:messagefees}
\end{table}

\paragraph{Running the botnet}
To have 1000 bots registered, 100 quota batches are needed, corresponding to 373000 sats.
Considering the fees for the batch transactions (45400 sats), this sums up to 418400 sats (0.004184 tBTC), which is then the amount required to register 1000 bots.
To fund the same number of bots, assuming an initial funding of 0.0001 tBTC each, and considering fees for the funding transactions (166000 sats), we have a total of 0.10166 tBTC.
This means that 0.1 tBTC (our estimated lower bound) are enough to register and fund 1000 bots per day.

For what concerns daily operations, assuming a specific behaviour is hard, as \cc communications for real botnets can be very diverse.
As such we will focus on the number of bytes that can be sent per day by a 1000-bot botnet, assuming it is funded with 0.1 tBTC per day.
To simplify things, we assume 1 sat is needed to send 1 byte of data.
This way, 0.1 tBTC is enough to send around 10MB per day, which translates to 10kB per bot in our example, which is likely to be insufficient for a modern botnet, according to available statistics \cite{correia2012statistical}.

However, by analyzing the Testnet blockchain, it is easy to see that a solo miner could obtain an average budget of as much as 4 tBTC per day, which would allow the botmaster to run, for instance, a spamming botnet, or to use this channel as a component of a larger hybrid botnet.

\subsection{Architecture analysis}
\paragraph{Testnet}
Despite being a testing network, Testnet is a very solid blockchain, as it constitutes a fundamental component of the Bitcoin ecosystem. In fact, it allows developers to test changes to the protocol and new applications without wasting money or messing the real chain.
Specifically, being released in 2012, the current version of the network (Testnet3) is one of the longest-running blockchains in the wild.
Although a new version might be introduced, this would affect a lot of ongoing projects and protocol improvements development, making it unlikely to happen soon.
As such, Testnet is a very stable backbone for a botnet \cc system.

A possible drawback of leveraging Testnet for a botnet might be its reduced network size, as fewer nodes might ease detection. 
However, the botmaster could mitigate this by deploying more nodes.

\paragraph{Faucets}
Faucets are a vital service for Testnet, as they allow developers to easily obtain the coins they need tu run their tests.
In their absence, developers would need to run a miner, making their job both harder and more expensive. 
As such, it is unlikely that such services will cease to work.

\paragraph{Bandwidth}
Despite the use of non-standard transactions in our protocol allows transmitting bigger amounts of data, 
message size is still limited compared to the traditional client-server model.
However, this system gains in terms of robustness, as communications are very hard to disrupt.

Given the above, it is possible that a real botnet would adopt a hybrid approach, with commands and responses happening on the blockchain, and larger data transmission being sent to a server, whose address changes periodically and gets updated via transactions.

\subsection{Security}

\paragraph{Stealthiness}
As communications happen via transactions, botnet messages will be permanently stored on the blockchain, creating an accessible evidence of past botnet activities and facilitating their analysis.
Furthermore, the use of non-standard transactions makes it easier to recognize botnet messages.
To mitigate this risk, the botmaster can limit their usage to only a part of the communications, trying to make other messages more similar to regular transactions.

\paragraph{Encryption}
All communications in our protocol are encrypted.
However, if a bot is compromised, the adversary can learn both the botmaster public key and the botnet key, enabling the monitoring of all the messages coming from the botmaster.
While this can help fighting the botnet activities, it does not prevent other bots from receiving and executing commands, thus being irrelevant to their operation.

To prevent this risk, the botmaster could encrypt and send messages individually for each bot.
This would make the protocol more expensive and less scalable but it might still be feasible if the botmaster were able to obtain coins at a fast rate.

\paragraph{Shared Account}
In case a bot is compromised an adversary can also learn the private key of the shared account and try to drain all the funds, preventing new bots from registering.

A possible solution for the botmaster would be to employ a backup registration system, such as an external channel where new bots can post their encrypted registration message. 
To avoid disruption, the botmaster can regularly change it and communicate the updated info via transaction\footnote{Note that bots are able to receive messages from the botmaster regardless of their registration status}.

Another way the adversary can steal funds is to register fake bots to get the corresponding coins sent by the botmaster.
This would increase the cost of the botnet and possibly make it infeasible to sustain.
The botmaster, however, can monitor and test bots to detect and ban misbehaving ones. 
As an additional precaution, the botmaster could initially send a smaller amount of coins, and only send more if the bot behaves as expected.

Another issue, related to the shared account, is that it allows to compute the size of the botnet in terms of spent quotas.
To mitigate this risk, the botmaster could periodically spend quotas at a random rate.
Although this would make the system slightly more expensive, it would effectively conceal the real number of bot registrations.

\paragraph{Countermeasures}
As mentioned above, 
the non-standard nature of the transactions used in our protocol allows to detect many of the botnet messages.
Additionally, if a bot is compromised, it is possible to monitor and decrypt all messages from the botmaster.
Furthermore, new bots can be prevented (or at least hindered) from registering.

Nonetheless, blocking botnet communications is hard as they are embedded into valid transactions.
If a botnet is detected, messages coming from the botmaster could be prevented from spreading.
However, this would be in sheer contrast with the anti-censorship principle at the base of the Bitcoin blockchain.

The most effective way to limit botnet communications would be to disallow non-standard transactions.
However, it is unclear how this would affect the regular operations of Bitcoin developers.

\section{Conclusion and Future Work}
\label{sec:conclusion}
In this paper, we showed how it is possible to implement a bidirectional encrypted \cc communication system on top of Bitcoin Testnet, which is both practical and economically affordable. We described a viable protocol that allows to register, fund, and control bots. Communications between bots and botmaster are encrypted and allow exchanging large amounts of data, enabling advanced functionalities, such as outsourcing bots code to the blockchain.
According to our estimates and experimental results, this system could be used in real life to run a small spamming botnet or as a component for larger hybrid botnet architectures.

This should call for an effort in either limiting the possibility of misusing Bitcoin Testnet for malicious purposes or devising appropriate countermeasures.

Future work includes a characterization of the communication patterns should be done to help designing effective detection mechanisms, as well as an analysis of strengths and weaknesses of this kind of botnet protocols, along with a study of valid alternatives.
Finally, an estimation of the impact that such malicious activities might have on the network could help to evaluate undesired side effects.

\bibliographystyle{plain}
\bibliography{ms}

\end{document}